\documentclass[twocolumn,aps,prl,amsmath,amssymb,preprintnumbers]{revtex4-1}
\usepackage{graphicx, color}
\usepackage[normalem]{ulem}

\begin{document}


\title{Physical implications of a fundamental period of time}

\author{Garrett Wendel}
\email{gmw5164@psu.edu}
\affiliation{Department of Physics, The Pennsylvania State
  University, 104  Davey Lab, University Park, PA 16802, USA}

\author{Luis Mart\'{\i}nez}
\email{lxm471@psu.edu}
\affiliation{Department of Physics, The Pennsylvania State
  University, 104  Davey Lab, University Park, PA 16802, USA}

\author{Martin Bojowald} 
\email{bojowald@gravity.psu.edu}
\affiliation{Department of Physics, The Pennsylvania State
  University, 104  Davey Lab, University Park, PA 16802, USA}

\begin{abstract}
  If time is described by a fundamental process rather than a coordinate, it
  interacts with any physical system that evolves in time. The resulting
  dynamics is shown here to be consistent provided the fundamental period of
  the time system is sufficiently small. A strong upper bound $T_{\rm C}<
  10^{-33}{\rm s}$ of the fundamental period of time, several orders of
  magnitude below any direct time measurement, is obtained from bounds on
  dynamical variations of the period of a system evolving in time.
\end{abstract}

\maketitle

Dimensional arguments are often used to suggest that time has a fundamental
period, given by the Planck time $t_{\rm P}=\sqrt{\hbar G/c^5}= 5.39 \times
10^{-44} {\rm s}$ using the speed of light $c$, Newton's constant $G$, and
Planck's constant $\hbar$. Resolving this time scale is far beyond currently
available technology. Nevertheless, it may be possible to obtain indirect
information about physics near this scale, much like Brownian motion helped to
confirm the atomic nature of matter using light microscopy, able to resolve
only distance scales much larger than the atomic size. In order to devise
indirect measurements, a detailed physical model must be available to derive
effects that could magnify the sensitivity of a direct measurement. Here, we
show that quantum mechanics of a physical model of time, described not as a
monotonic external parameter but rather as a dynamical and oscillating
variable that can model a physical clock, reveals a surprising magnification
effect. The resulting upper bound on a potential fundamental period of time is
about ten orders of magnitude above the Planck time, but much closer than any
direct measurement could provide.

Formulating quantum mechanics with a physical, oscillating time variable may
at first sight seem in conflict with the requirement of unitarity, which
implies that the evolution operator between two states, $\psi(0)$ and
$\psi(t)$, is given by $\hat{U}(t)=\exp(-i\hat{H}t/\hbar)$ using the
self-adjoint Hamiltonian $\hat{H}$ of the evolving system. If this condition
must be maintained for all $t$, it is impossible to make sense of a dynamical,
oscillating time variable which turns back to its initial value after each
clock cycle while the system does not, in general, evolve back. Moreover, even
during phases in which the expectation value of a physical, and therefore
quantum, time variable changes monotonically, the variable should be subject
to quantum fluctuations which do not have a preferred direction. These
problems are especially acute in quantum gravity and quantum cosmology, two
fields which aim to quantize generally relativistic systems in which there is
no absolute time \cite{KucharTime,IshamTime,AndersonTime}.

A proposal to formulate a meaningful notion of physical time goes back to
an investigation by Dirac \cite{GenHamDyn1} that analyzed general properties
of quantum constrained systems relevant for generally relativistic
systems. Dirac briefly suggested a construction, now called
deparameterization, which, with hindsight, can be interpreted as a solution to
the problem of quantum fluctuations of a physical time variable by showing
that physical time requires constrained dynamics: If both time and the system
of interest are quantized in an extended model that includes all relevant
degrees of freedom, the energies of the time variable and the system have to
be exactly balanced. Otherwise, a non-zero net energy would imply
non-trivial evolution of the extended model in an external absolute time
parameter, violating the assumption that time is described by an internal
degree of freedom. 

A specific example from cosmology is the
Friedmann equation
\begin{equation}
 \left(\frac{1}{a}\frac{{\rm d}a}{{\rm d}t}\right)^2= \frac{8\pi G}{3c^2}\rho
\end{equation}
for the scale factor $a>0$, with the energy density $\rho$ of matter. In
canonical form, this equation can be rewritten as an energy-balancing
constraint
\begin{equation}  \label{C}
C= - H_{\rm matter}(V)+\frac{6\pi G}{c^2} Vp_V^2=0
\end{equation}
where $V=a^3$ is the spatial volume and $p_V=-c^2\dot{a}/(4\pi G a)$ its
momentum. By virtue of the constraint, the matter energy
$H_{\rm matter}(V)=V\rho$ always equals the gravitational contribution $6\pi
G Vp_V^2/c^2$.

A common matter system in cosmological models is a scalar degree of freedom
$\phi$ with momentum $p_{\phi}$ and energy density
\begin{equation}
 \rho=\frac{1}{2}\frac{p_{\phi}^2}{V^2}+ W(\phi)
\end{equation}
where $W(\phi)$ is the scalar potential, such as
$W(\phi)=\frac{1}{2}m^2\phi^2$ for a scalar of mass $m$. It is then possible
to describe the expansion of the universe by a function $V(\phi)$ that
determines the volume with reference to the value of the scalar, rather than
using a time coordinate not described by a physical subsystem. To derive
$V(\phi)$ classically, one first writes Hamilton's equations of motion by
interpreting the constraint $C$ as the total Hamiltonian, such as ${\rm
  d}\phi/{\rm d}\epsilon= \partial C/\partial p_{\phi}= p_{\phi}/V$, and then
eliminates the auxiliary parameter $\epsilon$ from the solutions
$\phi(\epsilon)$ and $V(\epsilon)$, also using solutions for the momenta.
(Systematic expansions that do not require intermediate $\epsilon$-dependent
functions have been derived in
\cite{PartialCompleteObs,PartialCompleteObsII}).

Deparameterization, in cosmological models following the constructions of
\cite{Blyth}, assumes that the scalar used as time is massless and without
self-interactions, $W(\phi)=0$. Hamilton's equations derived from the
constraint $C$ then imply that $p_{\phi}$ is conserved, while the rate of
change of $\phi$, as just derived, is proportional to $p_{\phi}$. As long as
$p_{\phi}\not=0$, such that there is in fact energy in the time variable,
$\phi$ is always monotonic in $\epsilon$. The assumption of zero scalar
potential therefore does not allow one to describe oscillating clocks, but it
shows how standard evolution as in quantum mechanics can formally be
recovered: If we solve the constraint for $p_{\phi}$, we can quantize the
resulting equation $-p_{\phi}= \sqrt{12\pi G} V|p_V|/c$ by a Schr\"odinger
equation
\begin{equation} \label{Schroed}
 i\hbar \frac{\partial\psi}{\partial\phi}= \frac{\sqrt{3\pi G}}{c}
   \left(\hat{V}|\hat{p}_V|+|\hat{p}_V|\hat{V}\right)\psi\,.
\end{equation}
On solutions of the constraint, therefore, quantum fluctuations of $\phi$ do
not present an obstacle to unitary evolution because they are no longer
independent of the system degrees of freedom
\cite{EffTime,EffTimeLong,EffTimeCosmo,AlgebraicTime}.

However, deparameterization, in spite of its widespread use in quantum
cosmology, is not a realistic description of a fundamental process underlying
our measurements of time because it would require fine-tuned interactions that
prevent one variable from oscillating. The description of an {\em oscillating
  motion} would require the presence of a background time, such as the gauge
parameter $\epsilon$ used above. Referring to such a monotonic background
parameter might seem to render our logic circular. However, our construction
will make use of a more general definition of an {\em oscillating variable} as
one that enters a basic Hamiltonian with a standard kinetic energy and a mass
term or some (self-)interaction potential that is unbounded from above. If one
were to solve such a system in a background time, one would obtain oscillating
motion, but we will require only the stated condition on the generic
functional dependence of a Hamiltonian.  

While monotonic readings can be constructed in specifically designed clocks or
calendars, accounting for the fine-tuning required for a monotonic time
variable to emerge, they do not refer to fundamental variables that would
appear in a basic Hamiltonian with some potential or interaction term. Our
model below will, in fact, describe a construction that shows how a monotonic
time variable ($\tau$) can emerge from an oscillating fundamental variable
($\phi$).  For instance, if we include a mass term in the cosmological model,
$W(\phi)=\frac{1}{2}m^2\phi^2$, $\phi$ is an oscillating variable, and
$p_{\phi}$ is no longer constant on solutions of the constraint (\ref{C}). The
time variable $\phi$ has turning points whenever $p_{\phi}$ equals zero. 

A procedure to formulate quantum evolution with respect to such an oscillating
time variable has only recently been given \cite{Gribov}. We illustrate and
evaluate this procedure for a more familiar system from quantum mechanics
rather than quantum cosmology, using a constraint
\begin{equation} \label{Cp}
C'=p_{\phi}^2+\lambda^2\phi^2- H(x,p)^2
\end{equation}
where $H(x,p)$ is the Hamiltonian of a standard system such as the
harmonic oscillator. For $\lambda=0$, we can use deparameterization, such that
the quantized solution $-p_{\phi}= H(x,p)$ of $C'=0$ is equivalent to the
Schr\"odinger equation of quantum mechanics. For $\lambda\not=0$,
\begin{equation}\label{pphi}
-p_{\phi}=\sqrt{H^2-\lambda^2\phi^2}
\end{equation} 
is set equal to a time-dependent
Hamiltonian which becomes problematic if we try to interpret $\phi$ as a
global time that can take any real number because $p_{\phi}$ would not always
be real. 

In order to address this problem, one first constructs a global time variable,
$\tau$, such that $\phi(\tau)=\pm\tau+A$ with constant $A$ is linear in
$\tau$, with unit rate, between any two turning points of $\phi$. Different
phases of $\phi$, separated by turning points, are related by choosing
constants $A$ in each phase such that $\phi(\tau)$ is continuous with
alternating ${\rm d}\phi/{\rm d}\tau=\pm1$. Without changing the dynamics,
$\tau$ then provides a global, monotonically increasing time parameter that
unravels the motion of $\phi$, just like standard clocks unfold the circular
motion of the minute hand by moving forward the hour hand after one minute
cycle has been completed. (A similar procedure has been applied to a related
case in which non-monotonic behavior is a consequence of non-trivial topology
rather than turning points \cite{DiracChaos,DiracChaos2}.) In the example of
$C'$, (\ref{Cp}), we use
\begin{equation}
\phi(\tau)=\left\{\begin{array}{cl} 
      (4n+2)\phi_{\rm t}-\tau & \mbox{if }4n+1 \leq \tau/\phi_{\rm t} \leq
      4n+3 \\ 
      \tau -4n\phi_{\rm t} & \mbox{otherwise} \end{array}\right. 
\end{equation}
where $\phi_{\rm t}=H/\lambda$ characterizes turning points
($p_{\phi}=0$). The integer
\begin{equation} \label{n}
    n =\left \lfloor \frac{1+\tau/\phi_{\rm t}}{4} \right \rfloor
\end{equation} 
determines the number of cycles of $\phi$ the fundamental clock goes through
between time $0$ and time $\tau$. Evaluating (\ref{pphi}) in $\phi(\tau)$
always gives a real number, for any $\tau$.

In the second step, we formulate quantum evolution with respect to $\tau$ by
concatenating evolution operators for monotonic phases of $\phi$. The
Schr\"odinger equation implied by (\ref{pphi}) can be solved in the energy
eigenbasis $\psi_k$ of the system with Hamiltonian $\hat{H}$ with energy
eigenvalues $E_k$: $\psi_k(q,\phi) = \psi_k(q,0) \exp(i\Theta_k(\phi))$ with
the phase
\begin{equation} \label{Theta}
 \Theta_k(\phi)= -\frac{1}{2 \lambda \hbar} \left(
            \lambda \phi
            \sqrt{E_k^2-\lambda^2\phi^2}+E_k^2
            \arcsin\left(\frac{\lambda\phi}{E_k}\right)\right)\,.
\end{equation}
This phase is real only for $\phi$ between its turning points, and therfore,
as expected, $\phi$ does not provide global evolution. However,
$\Theta_k(\phi(\tau))$ is always real and implies global evolution with
respect to $\tau$. Because ${\rm d}\phi/{\rm d}\tau=\pm 1$ is not constant,
however, we should alternate the sign of $\Theta_k(\phi(\tau))$ in order to
describe evolution with respect to $\tau$ instead of $\phi$. (The correct
equation is a slight modification of (\ref{Schroed}), changing
$i\hbar\partial\psi/\partial\phi=\hat{H}\psi$ to
$i\hbar\partial\psi/\partial\tau= ({\rm d}\phi/{\rm d}\tau)\hat{H}\psi$.)

The construction just described introduces well-defined, unitary evolution
with respect to $\tau$, implying a realistic model of time in which a system
evolves relative to an oscillating quantum clock. The clock and the system are
interacting through the energy-balance condition $C'=0$, (\ref{Cp}), which
implies rather complicated time-dependent Hamiltonians for
$\lambda\not=0$. Numerical simulations can however be performed and reveal
several interesting and surprising properties. In order to bring these out
most clearly, we now specify the system Hamiltonian to be given by the
harmonic oscillator, but the relevant features have been confirmed numerically
also for anharmonic and atomic systems.

\begin{figure}
    \centering
    \includegraphics[width=0.5\textwidth]{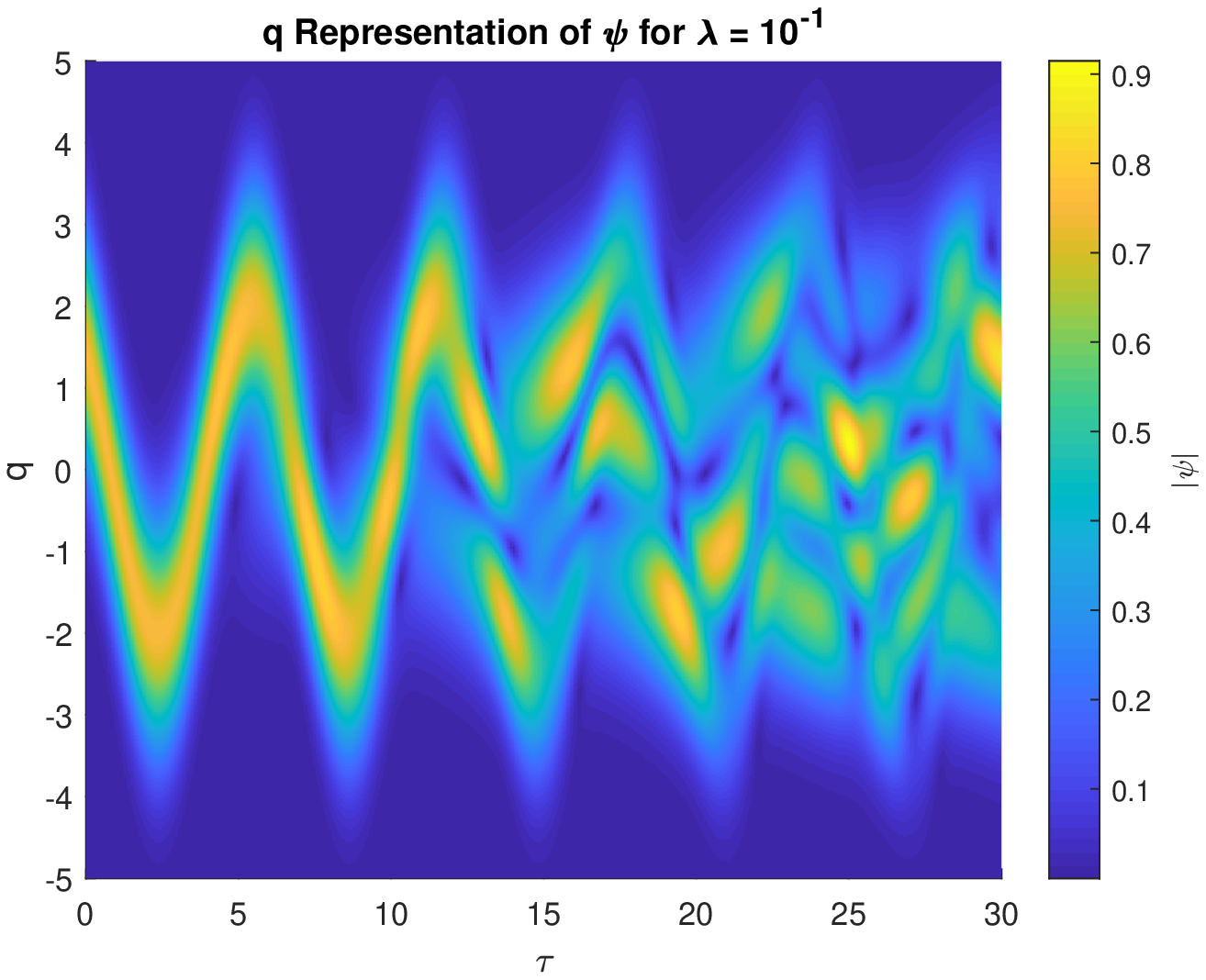}
\includegraphics[width=0.5\textwidth]{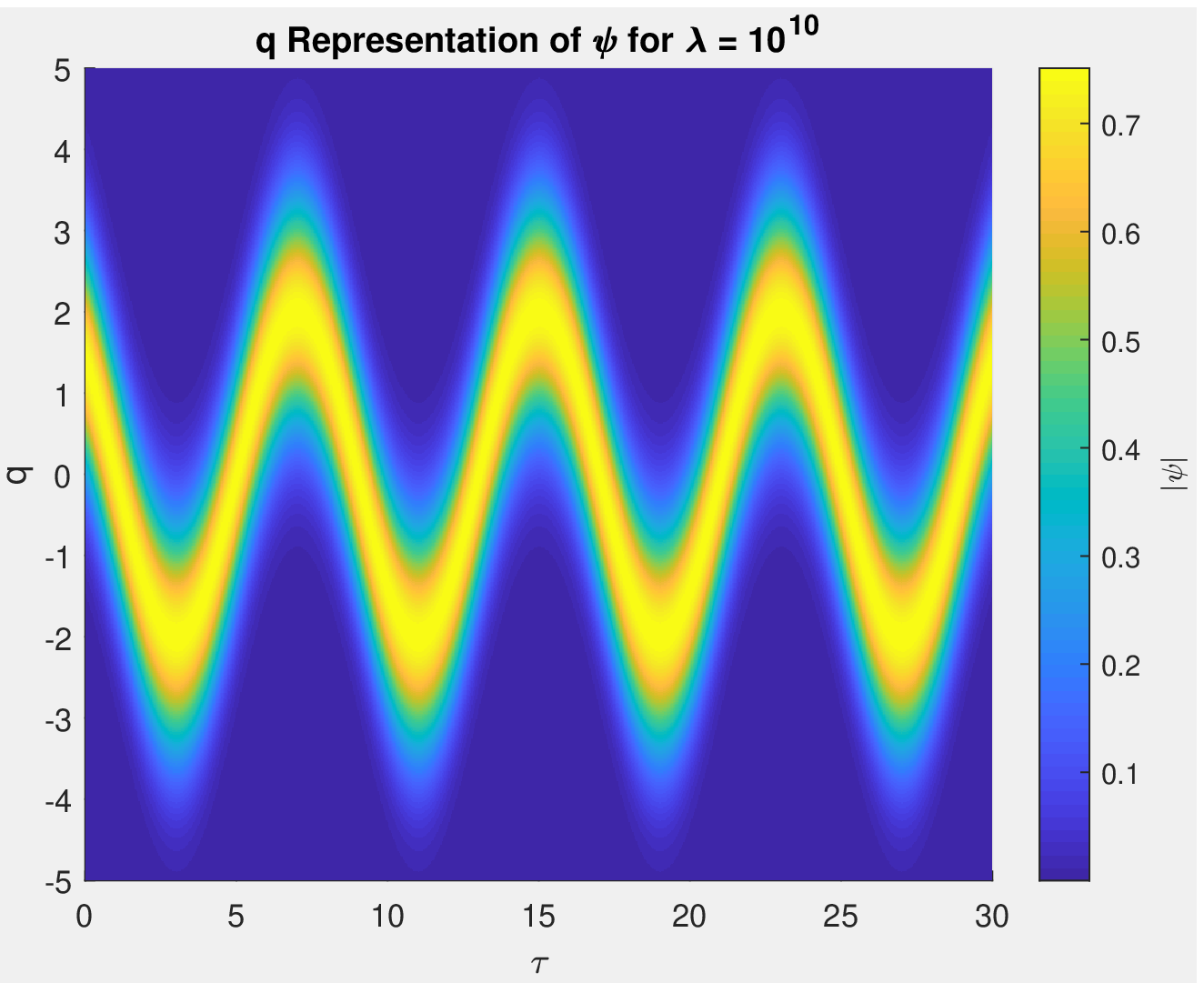}
    \caption{Density plots of the wave function for small $\lambda$
      (top) and large $\lambda$ (bottom), respectively,
      using a harmonic-oscillator Hamiltonian
      $\hat{H}=\frac{1}{2}(\hat{p}^2+\hat{q}^2)$ and a coherent initial
      state.  \label{Fig:WaveFunction}} 
\end{figure}

For the harmonic oscillator, we expect that the strong coherence of the
standard system, realized for $\lambda=0$, disappears for $\lambda\not=0$ in
which case the quadratic Hamiltonian is replaced by (\ref{pphi}). This
expectation is confirmed in Fig.~\ref{Fig:WaveFunction}. However, the same
figure shows that coherence remains intact for large $\lambda$, defined as
values of $\lambda$ such that the clock period $T_{\rm C}=4\phi_{\rm
  t}=4H/\lambda$ is much smaller than the system period. When the clock goes
through many cycles during a single system period, therefore, the dynamics is
almost indistinguishable from what is known from standard quantum mechanics. The
only visible difference is a system period rescaled by a factor of $\pi/4$,
which can always be absorbed in a redefinition of system parameters.

The rescaled period can be explained as follows: During each clock cycle,
while $\phi$ runs from $-\phi_{\rm t}=-E_k/\lambda$ to $\phi_{\rm t}$ and
back, the wave function accumulates a phase difference of
\begin{equation}
\Delta\Theta_k=2\left(\Theta_k(\phi_{\rm t})-\Theta_k(-\phi_{\rm t})\right)= -
\frac{\pi E_k^2}{\hbar} \frac{n}{\lambda}
\end{equation}
in each stationary state. Moreover, for large $\lambda$, the number of cycles,
$n$ given in (\ref{n}), divided by $\lambda$ can be approximated by
\begin{equation}
 \frac{n}{\lambda}= \frac{1}{4}\left \lfloor 1/\lambda+\tau/E_k \right \rfloor
 \approx \frac{\tau}{4E_k} \,. 
\end{equation}
Therefore, the phase accumulated over many cycles is approximately given by
\begin{equation}\label{DeltaTheta}
 \Delta\Theta_k\sim - \frac{\pi E_k\tau}{4\hbar}\,, 
\end{equation}
which is $\pi/4$ times the standard phase $-E_kt/\hbar$.

Our first result is therefore an unexpected revival of coherence for small
periods of a fundamental clock. Differences between deparameterization with a
monotonic time variable and the realistic implementation of a physical and
oscillating clock are tiny, providing justification for the deparameterization
procedure as a simplified mathematical method that is nevertheless able to
describe implications of an oscillating clock. Deparameterization, as
envisioned by Dirac, is therefore viable as a procedure that allows one to
understand qualitative features of relativistic quantum systems. (Other
aspects of the problem of time are still being studied, mainly related to
transforming observables obtained with different time or frame choices \cite{ReducedKasner,MultChoice,QuantumRef1,QuantumRef2,QuantumRef3,QuantumRef4}.)
However, our second result, to be described in the remainder of this letter,
shows that there are small effects of a physical, periodic clock that can be
relevant for sensitive observations.

\begin{figure}
    \centering
    \includegraphics[width=0.55\textwidth]{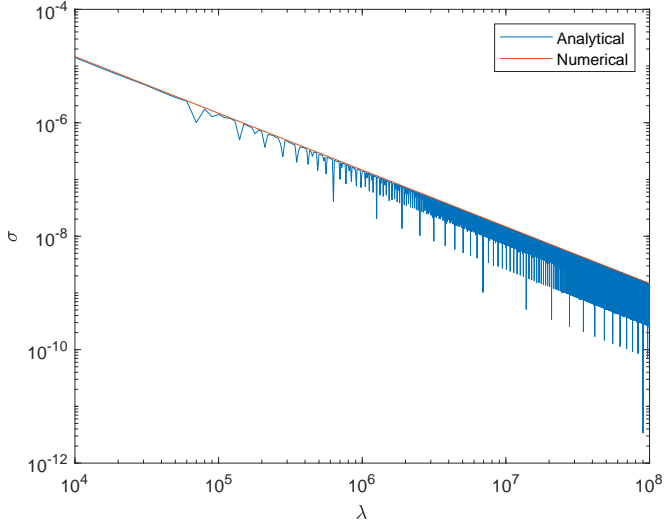}
    \caption{Relative standard deviation $\sigma$ of the system period over
      many system cycles as a function of $\lambda$. The analytical
      approximation (\ref{sigma}) agrees well with the upper limit of a
      numerical computation of many system periods, both confirming a
      $1/\lambda$-behavior of $\sigma$. \label{Fig:Sigma}}
\end{figure}

Additional deviations from the standard behavior can be uncovered by detailed
numerical analysis. In particular, because the phase (\ref{Theta}) is not
linear in $\tau$, the system does not go through its cycles in a uniform
manner, as would be the case with the standard linear phase $-E_kt/\hbar$ for
each stationary contribution. As a consequence, the distribution of system
periods taken over large evolution times has a non-zero standard deviation,
as shown in Fig.~\ref{Fig:Sigma}. Importantly, the plot shows a simple
$1/\lambda$-dependence of the standard deviation, which can be used in an
extrapolation to periods that would be too small for accurate numerical
evaluations.

Also the $1/\lambda$-behavior can be derived analytically for large
$\lambda$. We average the squared deviation of the phase (\ref{Theta}) from
the linear limit of $\lambda\to\infty$, $\Theta_k^{\infty}(\tau)=
\Delta\Theta_k$ given in (\ref{DeltaTheta}), over a half-cycle of $\phi$:
\begin{eqnarray} \label{sigma}
 \sigma^2 &=& \frac{1}{\phi_{\rm t}} \int_0^{\phi_{\rm t}}
   \left(\Theta_k(\phi(\tau))- \Theta_k^{\infty}(\tau)\right)^2{\rm
     d}\tau\nonumber\\
&=& \frac{E_k^4 (21\pi^2-1024/5)}{24^2\lambda^2\hbar^2}\,.
\end{eqnarray}
Therefore, the clock period $T_{\rm C}=4\phi_{\rm t}=4E_k/\lambda$ in a
stationary state is related to the system period $T_{\rm S}=2\pi\hbar/E_k$ by
\begin{equation} \label{TCTS}
 T_{\rm C}= \frac{48 \sigma T_{\rm S}}{\pi\sqrt{21\pi^2-1024/5}}\approx 9.7 \sigma
 T_{\rm S}\,.
\end{equation}

\begin{figure}
    \centering
    \includegraphics[width=0.5\textwidth]{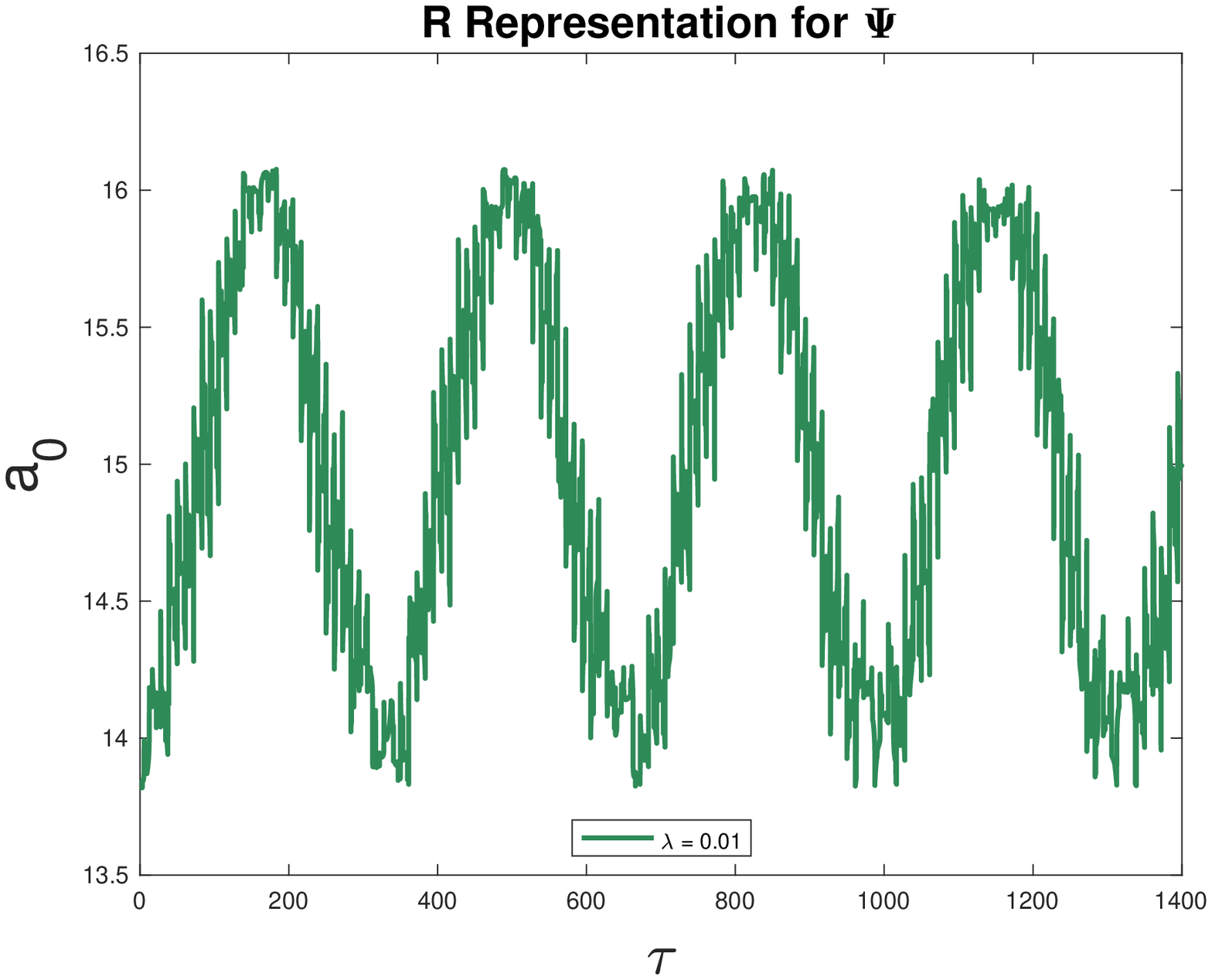}
\includegraphics[width=0.5\textwidth]{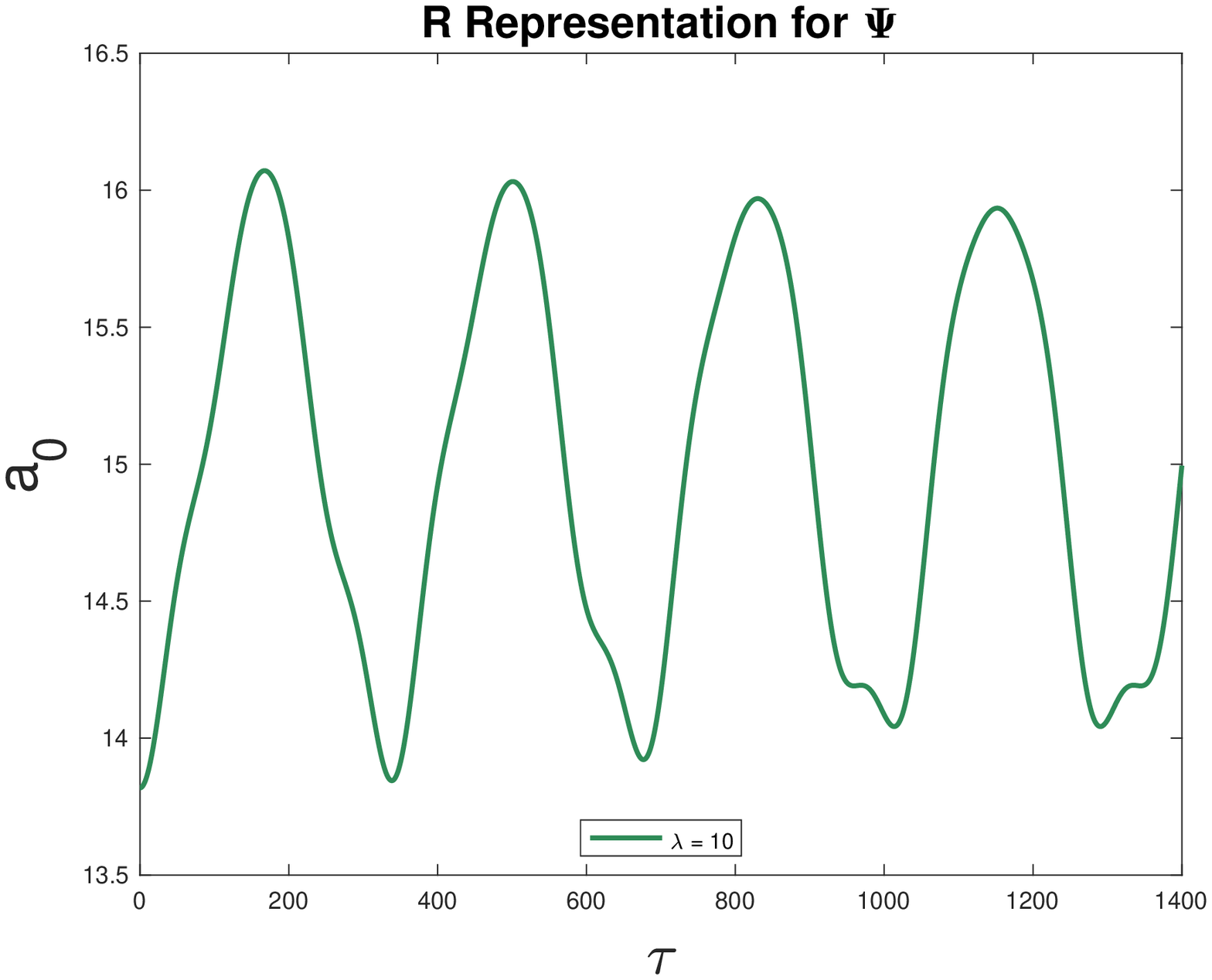}
\caption{Radius expectation values as multiples of the Bohr radius
  $a_0$ for small $\lambda$ (top) and large $\lambda$ (bottom),
  respectively, using a superposition of hydrogen eigenstates. The
  example of large $\lambda$ is visually indistinguishable from
  $\lambda=0$.  \label{Fig:Hydrogen}}
\end{figure}

Before we evaluate this result, we note that the qualitative behavior is
robust and does not depend much on the precise dynamics of the fundamental
clock. For a clock Hamiltonian other than $p_{\phi}^2+\lambda^2\phi^2$, the
phase $\Theta_k$ would be different, implying changes in the scaling factor of
$\pi/4$ in the system period and in the coefficients of (\ref{TCTS}). However,
results analogous to our specific equations would still be obtained. As long
as we are interested in an upper bound on the fundamental period of time,
therefore, the clock details do not matter. Considering systems other than a
harmonic oscillator, for the same fundamental clock, does not change our
results because we referred only to generic $E_k$ and $\psi_k$. They may be
more difficult to obtain for non-harmonic systems, but their specific form is
not required for our equations such as (\ref{sigma}). A non-harmonic example
is shown in Fig.~\ref{Fig:Hydrogen}.

Given the relative precision $\sigma$ of a time measurement, such as
$\sigma\approx 10^{-19}$ for recent atomic clocks \cite{LatticeClock} working
at a system period of $T_{\rm S}\approx 2{\rm fs}$ (corresponding to the wave
length $698{\rm nm}$ of the $^3P_0\rightarrow {}^1S_0$ transition of
Strontium), we therefore obtain the upper bound $T_{\rm C}< 10^{-33}{\rm s}$:
The measured precision could not be maintained if $T_{\rm C}$ were greater,
implying a non-uniform system period. This upper bound is about ten orders of
magnitude above the Planck time, but it is much smaller than could be achieved
with any direct measurement.

\noindent {\em Acknowledgements:}
This work was supported in part by NSF grant PHY-1912168. LM was supported by
a Gates Scholarship.


\begin{thebibliography}{10}

\bibitem{KucharTime}
K.~V. Kucha\v{r},  in {\em Proceedings of the 4th Canadian Conference on
  General Relativity and Relativistic Astrophysics}, edited by G. Kunstatter,
  D.~E. Vincent, and J.~G. Williams (World Scientific, Singapore, 1992).

\bibitem{IshamTime}
C.~J. Isham, {\em Integrable systems, quantum groups, and quantum field theory}
  (Kluwer, Dordrecht, 1993), pp.\ 157--287.

\bibitem{AndersonTime}
E. Anderson,  in {\em Classical and Quantum Gravity: Theory, Analysis and
  Applications}, edited by V.~R. Frignanni (Nova, New York, 2012).

\bibitem{GenHamDyn1}
P.~A.~M. Dirac, Can.\ J.\ Math. {\bf 2},  129  (1950).

\bibitem{PartialCompleteObs}
B. Dittrich, Gen.\ Rel.\ Grav. {\bf 39},  1891  (2007), gr-qc/0411013.

\bibitem{PartialCompleteObsII}
B. Dittrich, Class.\ Quant.\ Grav. {\bf 23},  6155  (2006), gr-qc/0507106.

\bibitem{Blyth}
W.~F. Blyth and C.~J. Isham, Phys.\ Rev.\ D {\bf 11},  768  (1975).

\bibitem{EffTime}
M. Bojowald, P.~A. H\"ohn, and A. Tsobanjan, Class.\ Quantum Grav. {\bf 28},
  035006  (2011), arXiv:1009.5953.

\bibitem{EffTimeLong}
M. Bojowald, P.~A. H\"ohn, and A. Tsobanjan, Phys.\ Rev.\ D {\bf 83},  125023
  (2011), arXiv:1011.3040.

\bibitem{EffTimeCosmo}
P.~A. H\"ohn, E. Kubalova, and A. Tsobanjan, Phys.\ Rev.\ D {\bf 86},  065014
  (2012), arXiv:1111.5193.

\bibitem{AlgebraicTime}
M. Bojowald and A. Tsobanjan,   arXiv:1906.04792.

\bibitem{Gribov}
M.~M. Amaral and M. Bojowald, Ann.\ Phys. {\bf 388C},  241  (2018),
  arXiv:1601.07477.

\bibitem{DiracChaos}
B. Dittrich, P.~A. Hoehn, T.~A. Koslowski, and M.~I. Nelson,
  arXiv:1508.01947.

\bibitem{DiracChaos2}
B. Dittrich, P.~A. Hoehn, T.~A. Koslowski, and M.~I. Nelson, Phys.\ Lett.\ B
  {\bf 769},  554  (2017), arXiv:1602.03237.

\bibitem{ReducedKasner}
P. Malkiewicz, Class.\ Quantum Grav. {\bf 29},  075008  (2012),
  arXiv:1105.6030.

\bibitem{MultChoice}
P. Malkiewicz, Class.\ Quantum Grav. {\bf 32},  135004  (2015),
  arXiv:1407.3457.

\bibitem{QuantumRef1}
F. Giacomini, A. Castro-Ruiz, and C. Brukner, Nat.\ Commun. {\bf 10},  494
  (2019), arXiv:1712.07207.

\bibitem{QuantumRef2}
A. Vanrietvelde, P.~A. Hoehn, F. Giacomini, and E. Castro-Ruiz, Quantum {\bf
  4}, 225 (2020),  arXiv:1809.00556.

\bibitem{QuantumRef3}
A. Vanrietvelde, P.~A. Hoehn, and F. Giacomini,   arXiv:1809.05093.

\bibitem{QuantumRef4}
P.~A. Hoehn, A.~R.~H. Smith, and M.~P.~E. Lock,   arXiv:1912.00033.

\bibitem{LatticeClock}
S.~L. Campbell {\it et~al.}, Science {\bf 358},  90  (2017).

\end{thebibliography}

\end{document}